# Biocompatible Writing of Data into DNA

Gary M. Skinner[1], Koen Visscher[1,2,3], and Masud Mansuripur[3]

[1]Department of Physics, The University of Arizona, Tucson, Arizona 85721, USA
[2]Departments of Molecular and Cellular Biology, The University of Arizona, Tucson, Arizona 85721, USA
[3]College of Optical Sciences, The University of Arizona, Tucson, Arizona 85721

**Abstract**. A simple DNA-based data storage scheme is demonstrated in which information is written using "addressing" oligonucleotides. In contrast to other methods that allow arbitrary code to be stored, the resulting DNA is suitable for downstream enzymatic and biological processing. This capability is crucial for DNA computers, and may allow for a diverse array of computational operations to be carried out using this DNA. Although here we use gel-based methods for information readout, we also propose more advanced methods involving protein/DNA complexes and atomic force microscopy/nano-pore schemes for data readout.

**1. INTRODUCTION**. DNA is attractive for storing digital information, partly due to its potentially ultra-high density[1]; in theory, 1 gram of DNA is capable of storing about the same amount of data as $10^{12}$ CD-ROMs.[2] DNA also offers the possibility of creating extremely durable information archives, e.g., by introducing the DNA into reproducing organisms, such as bacteria tolerant to radioactivity.[3] As the organism replicates its genome, the information is carried into the next generation. In combination with some form of selection pressure to reduce mutation rates, such information could be secured for thousands, perhaps millions, of years. Under appropriate conditions, even *in vitro* storage of DNA could be secure for hundreds of years. Another key attraction of DNA as a memory storage device is novel forms of computational problems that can be addressed using it. For example, in his seminal paper, Adleman demonstrated that DNA could be used to solve an instance of the Hamiltonian path or "Traveling Salesman" problem.[4] Such problems require large amounts of conventional computing time, even for problems of modest complexity. By making use of DNA's ability to rapidly search a large information space in parallel, DNA computing offers the possibility of solving these problems on a practical timescale. The DNA memory scheme presented here might further enable DNA computation, by allowing the DNA computer to be "programmed" in a faster, more efficient manner. This could save enormous time and expense over synthesizing each unique molecule one by one.

Previous groups have encoded meaningful information in DNA directly as a sequence of base pairs.[3,5] However, chemical DNA synthesis is slow and expensive, and a new molecule must be created each time new data is to be written. To overcome this limitation, methods have been developed that take advantage of the formation of complementary base pairs in DNA to generate molecules representing information. Notably, a 3-bit system has been developed by Shin and Pierce that allows for the encoding of up to eight distinct states in a DNA molecule, which is even rewritable.[6] This method has some important limitations, however, which may hinder its development into a fully-fledged DNA memory device. It was not possible to unambiguously read out the information using simple gel-based methods; in this case only the total number of memory bits that were "1" or "0" could be determined.[6] To resolve the ambiguities, a fluorescence-based system was included, using the quenching of different fluorophores at each memory location to determine the precise data content. However, this method is not practical for large capacity devices; a new fluorophore would be required for each new memory location, quickly exhausting all distinct fluorophore excitation/emission combinations available. A further limitation of this approach is that the resulting DNA construct is not easily manipulated by enzymatic or biological processes. Such a capability is critical, as the promise of DNA computers relies upon such manipulation of the encoded data to perform operations.[4] Specifically, Shin and Pierce's DNA construct is not linear, but resembles "frayed wires"[6] and, if introduced into a living organism, will not be faithfully replicated, and furthermore, it cannot be amplified using the polymerase chain reaction, as has been important in previous DNA computation demonstrations.[4] Although we do not claim that our method could be used directly in DNA computing, we hope that such ideas may be useful to those engaged in these endeavors.

In order to address the limitations of pre-existing DNA memory devices, we have developed an alternative DNA based memory that requires only simple methods to write information into DNA. Since the resulting molecules are linear double-stranded DNA, in principle this memory is compatible with a broad range of enzymatic and biological methods. Our method takes advantage of DNA recognition enzymes in order to distinguish between the "1" and "0" memory states. For this purpose we have chosen to use the restriction endonuclease EcoR1,[7] an enzyme that binds to the palindrome sequence G^AATTC, cutting both DNA strands at the site marked "^". We chose this enzyme as it is well known to possess high stringency for its cognate recognition site.[7] By engineering the memory states "0" and "1" so that in one state the enzyme can recognize the DNA, while in the other it cannot, we can unambiguously determine the information content by a simple EcoR1 digestion procedure. Similar methods for distinguishing DNA molecules have been routinely used for "DNA fingerprinting," and have been applied to solve certain criminal cases.[8] This approach is also somewhat similar to the method developed for the DNA "finite automation,"[9] developed by Benenson *et al*., in that a recognition enzyme is used to differentiate certain states of the system. An intriguing unidirectional motor has also been constructed from DNA using a sequence-recognizing nicking enzyme.[10] However, our device is different from these approaches in that we merely wish to create a generic memory device, enabling the easier encoding of a linear sequence of information directly into DNA.

Our device operates on the principle of using linear DNA as a "carrier" to specify a one-dimensional array of writeable memory sequences (see Figs.1 and 3). Each of these sequences is unique and the address space is large. For example, by using 24 bases for the address, there are potentially $4^{24}$ unique address sequences. One *memory* sequence on a carrier strand consists of two logical parts partitioned into three physical parts: two *addressing* sequences that flank one writeable *data* sequence. The writeable data sequence is composed of the recognition sequence (GAATTC) for the restriction endonuclease, EcoR1.[7] To write to a memory sequence, one oligonucleotide spanning the entire memory sequence and fully complementary to the address sequences is used. To write the "0", the oligonucleotide is further also fully complementary to the data sequence, whereas in the case of writing a "1" the oligonucleotide contains a 4-nucleotide mismatch, (G***TTAA***C) with the data sequence. Read-out of the memory now is based upon the ability of EcoR1 to recognize, bind, and cleave at its restriction site, the data sequence. In the case of a "1", however, the DNA structure is locally melted due to the mismatch, prohibiting EcoR1 recognition, binding, and cleavage. Only in the case when a "0" is written is the restriction site intact, i.e., double-stranded, and cleavable by EcoR1. The data sequence is flanked by two addressing sequences to reduce steric hindrance and allow sufficient physical space for EcoR1 enzymes to bind. Here we demonstrate that DNA cleavage by the EcoR1 enzyme provides a means to read out information from our memory device.

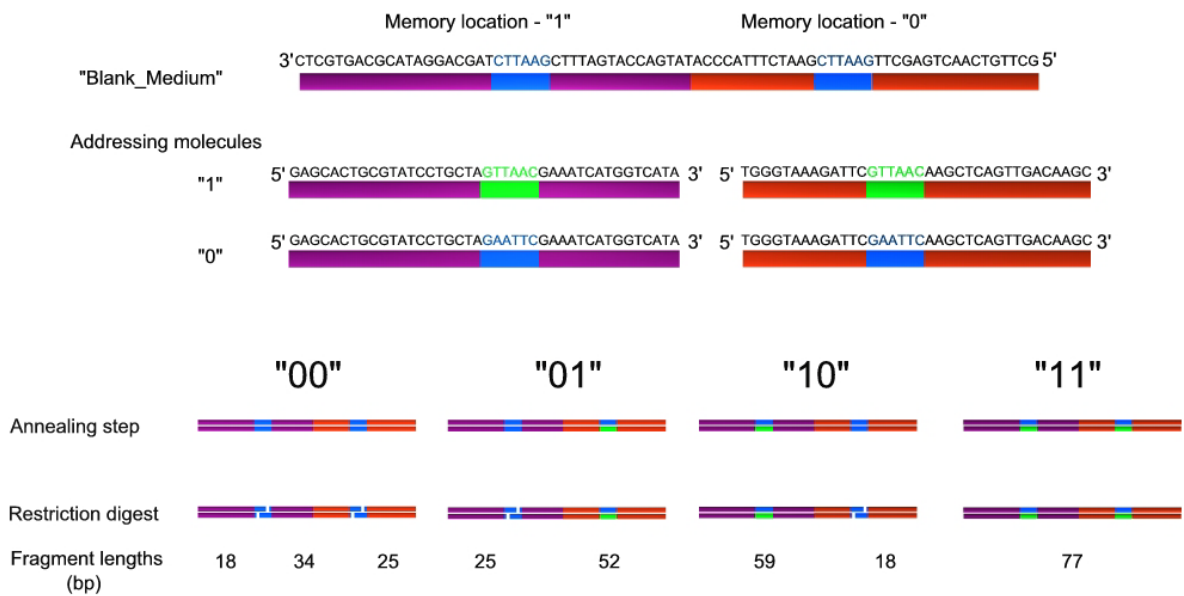

**Fig. 1**. Memory sequences used are shown at the top, with the EcoR1 sites (blue), and the addressing sequences for memory sequence 1 (magenta) and memory sequence 2 (red). In the addressing molecules, the binary digit "1" is represented by a partially mismatched EcoR1 site (green), while the binary digit "0" has the full complement to the EcoR1 site (blue). Shown are 2-bit sequences (00, 01, 10, 11) written into the DNA. Prior to annealing, the carrier and specific addressing molecules are added to the reaction to achieve the desired data sequence after annealing. Shown are the expected DNA fragments resulting from EcoR1 digestion of the encoded DNA products.

**2. METHODS**. In order to create a 2-bit capacity DNA based device with four distinct states (00, 01, 10, 11), five chemically synthesized ssDNA oligonucleotides were ordered from *Integrated DNA Technologies* (see Fig.1 for sequences). The carrier molecule consists of 77 bases and contains two memory sequences. For each memory sequence on the carrier, two shorter ssDNA oligonucleotides complementary to the addressing sequences and encoding either a "0" or a "1" were synthesized. The "0" molecules contain the complement to the EcoR1 sites as the data sequence, whereas the "1" molecules have four mismatched bases within this field (GAATTC → G*TTAA*C).

To store information using these molecules, one need only mix together the carrier molecule with either of the addressing molecules for each memory sequence, depending on whether a "0" or a "1" is desired. In practice, this was done by setting up a 50 μl annealing reaction in 1× NEB buffer #2 (10 mM Tris-HCl pH 7.9 @ 25°C, 50 mM NaCl, 10 mM $MgCl_2$, 1 mM dithiothreitol), with each ssDNA molecule present at 3 μM. The reaction was incubated at 95°C to melt all base-pair interactions, and then slowly cooled to room temperature over several hours. Following this step, the information was then stored within the molecules.

Subsequent readout of the stored digital information was achieved by performing a restriction digest with the enzyme EcoR1 (*New England Biolabs*). If a "0" was written at any given memory location, a completely base-paired cognate EcoR1 site is formed that can be recognized and cleaved by the enzyme. However, if a "1" was written, the four mismatched bases prevent the EcoR1 from recognizing and cleaving the DNA. By subsequent gel electrophoresis of the digestion products it is now possible to unambiguously determine the information encoded. In practice, 10 μl of the encoded product DNA was added to 40 μl NEB buffer #2 plus 40 Units of EcoR1. These reactions were incubated for 48 hours at 37 °C to ensure complete digestion of all EcoR1 sites. The reaction products were then analyzed by polyacrylamide gel electrophoresis (PAGE) in Tris-borate EDTA buffer at 10V/cm, stained with ethidium bromide, and imaged by using an ultraviolet light source. A 10 bp DNA stepladder was run alongside the samples in order to determine fragment lengths on the gel; see Fig.2.

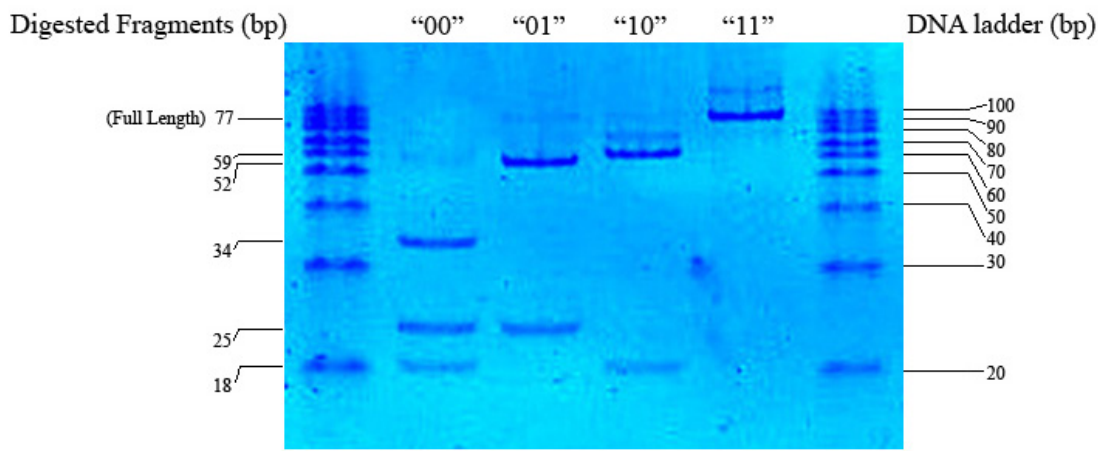

Fig. 2. Restriction digest with EcoR1 enzyme yields a distinct pattern of DNA fragments, which depends upon the specific EcoR1 sites that have been protected. In this way the four states of the 2-bit device (00, 01, 10, 11) can unambiguously be determined.

**3. RESULTS.** The molecules used in this demonstration were engineered so that the EcoR1 sites were not spaced evenly along the DNA molecule. This was done so that "01" could be distinguished from "10", as cutting at one or other of these sites produced differently sized fragments. The expected DNA fragment lengths for each state of the device are as follows (Fig.1). **"00"**: Since both data sequences are represented by fully base-paired cognate EcoR1, complete digestion yields three fragments, 18, 25, and 34 base-pairs in length. **"01"**: Only one site is cleavable yielding two fragments 25 and 52 base-pairs in length. **"10"**: Again, only one site it cleavable, but due to the spacing of EcoR1 sites, the fragments are 18 and 59 base pairs long. **"11"**: Since both EcoR1 sites are mismatched, no digestion can occur and the full-length 77 base-pair molecule is expected.

Following EcoR1 digestion, the resulting products were analyzed using a native polyacrylamide gel; see Fig.2. As can be seen, each state of the DNA memory device is easily distinguished from the other three and the expected DNA fragments are present as described above. Each of the four states "00", "01", "10" and "11" has been faithfully recorded into the DNA and can be unambiguously read using EcoR1, as expected.

**4. DISCUSSION**. The presented DNA memory device is easily scaled to much larger capacities; one need only increase the overall length of the carrier molecule to include additional memory sequences and synthesize the corresponding addressing molecules for them. It should be noted that these molecules must be made, in the first instance, by chemical synthesis. However, following these initial investments, any arbitrary data can be stored without the need for synthesis of further molecules, and further stocks of the carrier molecule could then be prepared by amplification using PCR reaction, followed by conversion to ssDNA using established methods.[11] In this way, our DNA memory device is a truly generic storage medium. We note that current DNA synthesis methods are limited to around 10,000 bases (e.g., gene synthesis by such companies as *GenScript*, http://www.genscript.com). However, we anticipate that longer sequences could be made by ligation of several synthetic genes, and perhaps even longer constructs will be possible as DNA synthesis technologies improve.

In this prototype device, the EcoRI sites were engineered so as to yield differently sized fragments upon digestion, e.g., to distinguish "01" from "10". As the capacity of such a memory device increases, this method for removing ambiguity will become increasingly impractical. In a larger device, the sites should be equally spaced along the molecule and the data stored as a palindrome within the DNA, i.e., all memory sequences duplicated and reflected around the center of the molecule; see Fig.(3a). With such a construct, all states will yield a unique pattern of digestion products. This has the added advantage of introducing redundancy into the memory, while decreasing the information density by only a factor of 2.

The resulting DNA memory is ready for use in further enzymatic manipulations; the fact that EcoR1 is able to digest the product efficiently already shows that it is a suitable substrate for at least this class of DNA processing enzymes. By sealing the single-stranded "nicks" in the DNA between adjacent memory locations, using ligase, it should be possible to amplify such DNA by PCR, and potentially to introduce it into living systems; such steps have been shown to be crucial for successful DNA information technology applications[3-5]. It would be true that only 50% of the PCR amplified DNA would be of the sequence desired, but by molecular cloning of these sequences into single-copy plasmids (e.g., *Novagen*'s pETcoco-1), the clones could be screened and only the desired sequence selected from resulting transformants.

We do not expect that restriction digestion and gel analysis will be the method of choice for read-out of the data in a practical application, especially when the capacity is increased. The gel method used here has two distinct limitations, first it is slow, and very long gels would have to be run to resolve the digestion fragments in larger devices. Furthermore, as the data is read in this fashion, the information is destroyed. However, we propose an alternative read-out strategy that would overcome these limitations. A mutant version of EcoR1 exists that has the ability to bind tightly to the cognate DNA site, but is unable to cut the molecule.[12] It simply remains bound, forming a protein "stud" on the DNA; such mutant EcoR1/DNA complexes studded along a length of λ-DNA have been imaged using atomic force microscopy (AFM).[12] It should be possible to use a similar method to decorate our memory device with EcoR1. A protein would bind to the cognate sites located at the "0" positions, while leaving gaps where the EcoR1 is unable to bind the mismatched sites that represent "1". The readout could be performed using AFM and subsequent image analysis, as was done with EcoR1 on λ-DNA.[12] However, a more exciting prospect is to feed the protein/DNA complex through a solid-state nanopore[13] forming a "nanoscale digital tape drive"; see Fig.3(b). We have previously shown, using the $\alpha$-hemolysin protein pore, that structural landmarks on ssDNA can be distinguished and potentially used to store information.[1] The pore diameter would be engineered to yield a large current blockade differential when EcoR1/DNA complexes pass through as opposed to when naked DNA strands pass through. The palindrome nature of the encoded data means that there is no need to control the orientation of the DNA molecule as it enters the nanopore. Thus, the data could be read as a sequence of blockade events, and the intact protein/DNA complex could then be returned to a storage area to be accessed later.

By having a robust, simple means to encode a sequence of digital information into DNA in a form that is suitable for a wide range of further manipulations, the goal of practical, large-scale DNA storage may be a step closer to reality.

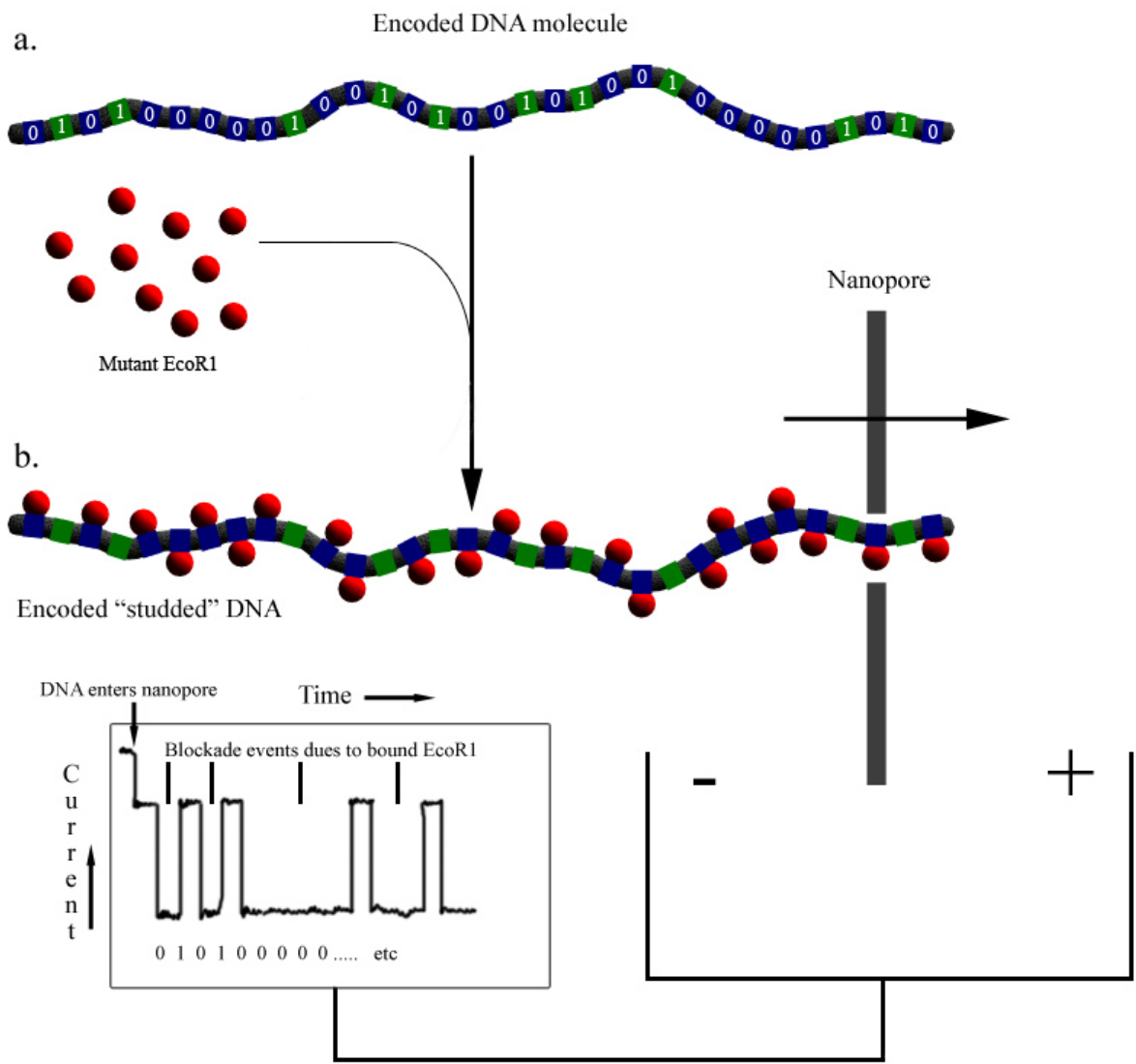

**Fig. 3**. Advanced readout system. (a) The encoded DNA molecule is incubated with the mutant form of EcoR1 that is able to bind tightly to its correct DNA site but is not able to cleave the DNA.[8] (b) The resulting "studded" DNA is then fed through a solid-state nanopore, driven by an applied voltage. The current through the nanopore is measured as a function of time and discrete additional blockade events would be observed (simulated here) each time an EcoR1/DNA complex passes through the nanopore. This current versus time data could then be used to determine the digital sequence of the encoded DNA.

**Acknowledgements:** This work is supported by the Office of Naval Research MURI grant No. N00014-03-1-0793, G.M.S. is a fellow of the University of Arizona BIO5 Institute and also thanks the Jane Coffin Childs Memorial Fund for Medical Research (www.jccfund.org) for support.